\documentstyle[12pt,a4wide]{article}
\input epsf

\newcommand{\news}{\setcounter{equation}{0}}
\newcommand{\be}{\begin{equation}}
\newcommand{\ee}{\end{equation}}
\newcommand{\bea}{\begin{eqnarray}}
\newcommand{\eea}{\end{eqnarray}}
\newcommand{\bean}{\begin{eqnarray*}}

\newcommand{\eean}{\end{eqnarray*}}
\newcommand{\beq}{\begin{equation}}
\newcommand{\eeq}{\end{equation}}
\font\upright=cmu10 scaled\magstep1
\font\sans=cmss12
\newcommand{\ssf}{\sans}
\newcommand{\stroke}{\vrule height8pt width0.4pt depth-0.1pt}
\newcommand{\Z}{\hbox{\upright\rlap{\ssf Z}\kern 2.7pt {\ssf Z}}}

\newcommand{\C}{{\rlap{\rlap{C}\kern 3.8pt\stroke}\phantom{C}}}
\newcommand{\R}{\hbox{\upright\rlap{I}\kern 1.7pt R}}
\newcommand{\CP}{\C{\upright\rlap{I}\kern 1.7pt P}}
\newcommand{\identity}{{\upright\rlap{1}\kern 2.0pt 1}}
\newcommand{\qd}{(q-q^{-1})}
\newcommand{\maple}{{\scriptsize MAPLE\ }}
\newcommand{\tK}{\widetilde{K}}
\newcommand{\Iim}{\mbox{Im}}
\newcommand{\Ad}{\mbox{Ad}}
\newcommand{\tT}{\widetilde{T}}
\newcommand{\tG}{\widetilde{G}}
\newcommand{\iv}{^{-1}}
\newcommand{\HH}{\mbox{\hbox{\upright\rlap{I}\kern 1.7pt H}}}
\newcommand{\Q}{{\rlap{\rlap{Q}\kern 3.0pt\stroke}\phantom{Q}}}
\newcommand{\one}{\bf 1}
\newcommand{\two}{\bf 2}
\newcommand{\thr}{\bf 3}
\newcommand{\fr}{\bf 4}
\newcommand{\fv}{\bf 5}
\newcommand{\six}{\bf 6}
\newcommand{\sev}{\bf 7}
\newcommand{\twp}{\bf 2'}
\newcommand{\thp}{\bf 3'}
\newcommand{\frp}{\bf 4'}

\begin{document}
\pagestyle{plain}
\title{\vskip -70pt
\begin{flushright}
\end{flushright}
\vskip 20pt
{\bf \Large \bf Symmetric Instantons and Skyrme Fields}
 \vskip 10pt
}
\author{Michael A. Singer$^{\ \dagger}$
and 
Paul M. Sutcliffe$^{\ \ddagger}$\\[10pt]
{\normalsize
$\dagger$ {\sl Department of Mathematics and Statistics,}}\\
{\normalsize {\sl The University of Edinburgh, Edinburgh, UK. }}\\
{\normalsize {\sl Email michael@ed.ac.uk}}\\[10pt]
{\normalsize 
$\ddagger$ {\sl
Institute of Mathematics,
University of Kent at Canterbury,}}\\
{\normalsize {\sl Canterbury CT2 7NZ, England.}}\\
{\normalsize {\sl Email P.M.Sutcliffe@ukc.ac.uk}}\\[10pt]
}

\date{January 1999}
\maketitle

\begin{abstract}
By explicit construction of the ADHM data, we prove the existence
of a charge seven instanton with icosahedral symmetry. By computing
the holonomy of this instanton we obtain a Skyrme field which
approximates the minimal energy charge seven Skyrmion. 
We also present a one parameter family of tetrahedrally symmetric
instantons whose holonomy gives a family of Skyrme fields which
models a Skyrmion scattering process, where seven well-separated
Skyrmions collide to form the icosahedrally symmetric Skyrmion.

\end{abstract}
\newpage
\renewcommand{\thepage}{\arabic{page}}

\section{Introduction}
\news\ \ \ \ \ \
Skyrmions are a type of topological soliton in three-dimensional space which are
of interest to physicists, in that they are candidates for a solitonic description of
nuclei. Numerical simulations reveal that the minimal energy Skyrmions often have
a great deal of symmetry, and in particular the charge seven Skyrmion has icosahedral
symmetry\cite{BS2}.

One approach to the study of Skyrmions is the suggestion by Atiyah and Manton\cite{AM}
that low energy Skyrmions of charge $n$ may be approximated by computing the holonomy
of a charge $n$ instanton in Euclidean $\R^4$ along lines parallel to the Euclidean time-axis.
This proposal therefore predicts the existence of a charge seven instanton with icosahedral
symmetry. In this paper we verify that indeed such an instanton exists by presenting the
corresponding ADHM data. We then use this data to compute the holonomy of the instanton
and hence generate a Skyrme field of charge seven with icosahedral symmetry, which is
an approximation to the minimal energy Skyrmion.

Manton has proposed\cite{Ma7} that the low energy dynamics of $n$ Skyrmions may be approximated
by motion on a finite dimensional manifold of charge $n$ Skyrme fields, and
 one reasonable choice for such a manifold appears to be the moduli space of 
$n$-instantons\cite{AM}.  In such an approximation a scattering of seven Skyrmions
is described by a one-parameter family of 7-instantons. We present such a family
of instantons, obtained by imposing tetrahedral symmetry, and compute the associated Skyrme
fields. We display the baryon density isosurfaces for this scattering process, in which six Skyrmions
approach along the Cartesian axes a seventh Skyrmion at the origin. The Skyrmions merge to
form first a dodecahedron, then a cube, and finally the dual dodecahedron before again separating
along the Cartesian axes leaving a single Skyrmion remaining at the origin. 
Part of this scattering process, where the dodecahedron
deforms to its dual, is believed to be an important vibrational mode of the dodecahedral Skyrmion,
and these are of relevance when considering the quantization of Skyrmions\cite{BBT}.

\section{An icosahedral 7-instanton}
\news\ \ \ \ \ \
In this section we shall present the ADHM data for an icosahedrally
symmetric 7-instanton.  First, we briefly recall the ADHM
construction of instantons\cite{ADHM} and give an explanation of
symmetry in a gauge theory.

The ADHM data for an $SU(2)$ $n$-instanton consists of a matrix
\be
\hat M=\pmatrix{L\cr M}
\label{split}
\ee 
where $L$ is a row of $n$ quaternions and $M$ is a symmetric
$n\times n$ matrix of quaternions.  By the {\em ADHM constraints} we
shall mean the condition
\be\label{adhm1}
M^\dagger M \mbox{ is a real matrix},
\ee
where $M^\dagger$ denotes the quaternionic conjugate transpose
of $M$.

The first step in constructing the instanton from the ADHM data is to
form the matrix 
\be \label{eqdel} \Delta(x)=\pmatrix{L\cr
M-x\identity_n} \ee 
where $\identity_n$ denotes the $n\times n$
identity matrix and $x$ is the quaternion corresponding to a point in
$\R^4$ via $x=x_4+ix_1+jx_2+kx_3.$

The second step is then to find the $(n+1)$-component column vector $N(x)$, of unit length,
which solves the equation
\be
N(x)^\dagger \Delta(x)=0.
\label{construction}
\ee

The final step is to compute the gauge potential $A_\mu(x)$ from
$N(x)$ using the relation 
\be A_\mu(x)=N(x)^\dagger\partial_\mu N(x).
\label{afromn}
\ee This defines a pure quaternion which can then be regarded as an
element of $su(2)$ using the standard representation of the
quaternions in terms of the Pauli matrices.

In order that these steps make sense it is necessary that the ADHM
data satisfy an additional invertibility condition: that the columns of
$\Delta(x)$ span an $n$-dimensional quaternionic space, for every
$x$. Equivalently this condition can be expressed as:
\be\label{adhm2}
\Delta(x)^\dagger\Delta(x)\mbox{ is invertible for every }x.
\ee
In the absence of this condition ADHM data satisfying (\ref{adhm1})
only gives rise to a self-dual gauge field with singularities
(corresponding to the points $x$ where (\ref{adhm2}) fails).

\subsubsection{Symmetric instantons} 

It will be useful to start by considering the problem of symmetric
instantons in a general context. Thus let $G \subset SO(3)$ be a
subgroup of the rotation group of Euclidean $3$-space, and make it act
on $\R^4$ by rotations on $(x_1,x_2,x_3)$, leaving $x_4$ alone.  This
action has a very convenient description using quaternions. For
this, replace $G$ by the corresponding binary group $\tG\subset SU(2)$
(the double cover of $G$ obtained from the double cover $SU(2) \to
SO(3)$).  Now
\be \label{dcov}
SU(2) = Sp(1) = \{u \in \HH: uu^* = u^*u = 1\}
\ee
and $x \mapsto uxu^{-1}$ clearly preserves the $x_4$ (real) component
of $x$. It also acts on the imaginary part of $x$ by the rotation in
$SO(3)$ corresponding to $u$ in $SU(2)$. (In fact this gives a
{\em construction} of the double cover of $SO(3)$ by $SU(2)$.)  From
now on, whenever we speak of a subgroup of $SO(3)$ acting on $\R^4$,
we shall mean that it acts in the way we have just described.

In order to get a grip on the problem of imposing symmetry in a gauge
theory, let us introduce a rank-2 vector bundle $E$ over $\R^4$ and
assume that our gauge potential $A_\mu$ defines a connection on $E$.
If $\tG$ acts on $E$ then it makes good sense to require $A_\mu$ to be
invariant (or symmetric) under this action. In this case an action of
$\tG$ on $E$ consists of the data of a unitary map $\Omega(g,x): E_x \to
E_{gxg\iv}$ for each $g\in \tG$ and $x\in \R^4$, depending smoothly
on $x$, and satisfying $\Omega(1,x)=$ the identity, and
\be\label{act1}
\Omega(f,gxg\iv)\Omega(g,x) = \Omega(fg,x) \mbox{ for all }f,g,\in
\tG,\,x\in \R^4
\ee
both sides of this equation being maps $E_x \to E_{(fg)x(fg)\iv}$. 

It is important to note that one cannot assume that there exists a
gauge in which all the $\Omega$'s are equal to the identity.  What is
true, however, is the following.  Let $a$ be any fixed-point of the
action of $\tG$, i.e.\ any point on the $x_4$-axis. For such a point,
(\ref{act1}) reads  $\Omega(f,a)\Omega(g,a) = \Omega(fg,a)$ for all $f,g,\in
\tG$, giving a {\em representation} of $\tG$ on the fibre $E_a$. This
representation changes to an equivalent one under changes of gauge, so
if this representation is non-trivial, there cannot be any gauge with
the $\Omega$'s all equal to the identity.   

Conversely, given any complex $2$-dimensional representation $\rho$ of
$\tG$, we may define an action of $\tG$ on a  bundle over $\R^4$ by putting
\be\label{int}
\Omega(g,x) = \rho(g)
\ee
relative to some background gauge. Obviously this action restricts to
the representation $\rho$ on the fibre $E_a$. In fact, if $\tG$ is a
finite subgroup of $SU(2)$, {\em any} action of $\tG$ on $E$ is of
this form for some representation $\rho$. (This is so because $\R^4$
is topologically trivial and can be contracted to the fixed-point set
of the action.  In general the classification of actions on bundles is
more complicated.)

Given a bundle $E$ with an action $\Omega$, we can state the condition
for a gauge potential to be symmetric:
\be\label{psym}
A_\mu(gxg\iv) = \Omega(g,x)A_\mu(x)\Omega(g,x)\iv -
\partial_\mu\Omega(g,x)\Omega(g,x)\iv.
\ee
In this context, the $\Omega$'s are often referred to as `compensating
gauge transformations' a practice we shall occasionally follow in this
paper.
The  term $-\partial_\mu\Omega(g,x)\Omega(g,x)\iv$
vanishes if the action is in the standard form (\ref{int}). Notice
that if (\ref{psym}) is satisfied then the curvature (field-strength)
transforms like $F(gxg\iv) = \Omega(g,x)F(x)\Omega(g,x)\iv$ and for example
the action density $-\mbox{tr}(F_{\mu\nu}F^{\mu\nu})$ is a $G$-invariant
function on $\R^4$.\\

\noindent{\bf Technical Remark} Since the object of physical
significance, the gauge potential, takes its values in the adjoint
bundle of $E$, it is most natural to impose symmetries on this bundle.
Thus we suppose that $G$ acts on $\Ad(E)$, covering the action of $G$
on $\HH$ (so $g\in G$ gives a linear map $\Ad(E)_x \to \Ad(E)_{gxg\iv}$
for every $x$ in $\HH$). Then it does not necessarily follow that such
an action of $G$ on $\Ad(E)$ lifts to an action of $\tG$ on $E$.
Moreover this failure occurs in at least one physically interesting
example, the cubic instanton of \cite{LM}. 

To explain how this problem
can be understood, we remark that, as above, the action of $G$ on
$\Ad(E)$ restricts to give a representation of $G$ on $\Ad(E)_a$
for any point $a$ fixed by the action. Since $\Ad(E)_a$ may be
identified with the Lie algebra of $SU(2)$ and hence with $\R^3$, in
general the image of $G$ by this representation will be a subgroup of
$SO(3)$ isomorphic to some quotient group $F=G/H$ of $G$.  In general
the double cover $\widetilde{F}\subset SU(2)$ of $F$ acts on $E_a$ and
this extends, as in (\ref{int}) to an action on the whole of
$E$. However it is in general {\em not} the case that $\widetilde{F}$
is a quotient of the binary group $\tG$.  In the case of the cubic
instanton, $G$ is the rotation group of the cube, $H$ is the Klein
viergruppe (consisting of the identity and the half-turns about the
three coordinate axes) and $F$ is isomorphic to the dihedral group of
order $6$.  

In the general case, it still happens that a double cover $\widehat G$,
say, acts on $E$, but this need not be the binary group $\tG$. With
$F,\widetilde{F}$ as above, 
$$
\widehat G = \{(g,f) \in G\times \widetilde{F}: p(g) = q(f)\}.
$$
Here $p: G \to G/H= F$ is the map to the factor group and
$q:\widetilde{F}\to F$ is the double-cover.  The restriction to $\widehat
G$ of the projection $G\times \widetilde{F} \to G$ is then a $2:1$
map, as one can easily verify from the definitions.

If $G$ is the icosahedral group, this complication does
not arise owing to the fact that $G$ is then a simple group (being
isomorphic to the alternating group $A_5$) and $H$ must either be $1$
or $G$. 

\subsection{Symmetric ADHM data}

Returning to the ADHM description (\ref{eqdel}), this will be $\tG$-symmetric
if for every $g \in \tG$, we have `compensating gauge
transformations' 
\be \label{inv1}
\Delta(gxg\iv) = \pmatrix{\rho_\infty(g) & 0\cr 0 & U(g)\cr}\Delta(x)U(g)\iv
\ee
for every $g$ in $\tG$. The first matrix has been decomposed into
blocks corresponding to (\ref{split}) so $\rho_\infty$ is $1\times 1$ and
$U$ is $n\times n$. In order to preserve the shape of $\Delta(x)$,
$U(g)$ must be the product of a real orthogonal matrix with a unit
quaternion, while $\rho_\infty(g)$ can be any unit quaternion. (Recall that
the matrix $M$ is {\em symmetric}.) 

Considering the coefficient of $x$ on each side of (\ref{inv1}), we obtain
\be\label{inv3}
U(g) x U(g)\iv = g xg\iv\identity_n.
\ee
By taking $x=1,i,j,k$ one deduces that $U(g) = \rho_w(g)\cdot
g$ where $\rho_w(g)$ is real.  Hence ADHM data are $\tG$-invariant if
\be \label{inv4}
\rho_\infty(g)L\rho_w(g)\iv g\iv = L,\;\;
\rho_w(g)g M\rho_w(g)\iv g\iv = M,
\ee
where $\rho_w$ is a {\em real} $n$-dimensional representation of $\tG$ and
$\rho_\infty$ is a quaternionic $1$-dimensional representation of $\tG$.

The reason for the notation ($\rho_w,\rho_\infty$) is that
$\Delta(x)$ can be viewed more invariantly as an $\HH$-linear map $W \to
W\otimes \HH \oplus E_\infty$ where $W$ is an $n$-dimensional real
vector space and $E_\infty$ is the fibre at $\infty$ of the bundle
carrying our $SU(2)$ gauge potential. (The space $W$ can in turn be
identified with the zero-modes of the coupled Dirac operator, but we
shall not need this fact.) Here we are using the $SU(2)$-structure to
think of $E_\infty$ as a $1$-dimensional quaternionic vector space.

Now the ADHM construction is natural; if we have an action of $\tG$ on
$E$ (covering the action of $G$ by rotations as above), such that the
gauge potential is 
$G$-symmetric, then $W$ and $E_\infty$ automatically become
representation spaces for $\tG$. The above notation reflects the
origin of the representations $\rho_w$ and $\rho_\infty$.

The reason why the ADHM equations become tractable after symmetry is
imposed is simply that if $W$ is a sum of not too many irreducible
representations of $\tG$, then there will not be too many parameters
involved in the specification of $L$ and $M$ satisfying (\ref{inv4}).
Our next task, then, is to consider the irreducible representations of
$\tG$ and the construction of $\tG$-invariant maps between tensor
products of certain of these representations.

\subsection{A little representation theory}
\label{repthy}
The problem of choosing the representations $\rho_w$ and $\rho_\infty$
and of constructing the invariant matrices (\ref{inv4}) will now be
considered.  Let us first compare it with the analogous problem of
constructing symmetric Nahm data (and hence symmetric
monopoles \cite{HMM,HS2,HS1}).  The single most important difference
between these 
problems is that in the Nahm case one knows which representation (the
analogue of $\rho_w$) is going to arise.  That is because one knows
that the Nahm data form the irreducible $n$-dimensional representation
of $SU(2)$ at the end-points, and this $SU(2)$ really does correspond
to the rotation group of $\R^3$.  Thus in the cited work on
symmetric monopoles, the approach was to understand
explicitly how the standard representations of $SU(2)$ decompose under
the action of $\tG$, making use of the invariants corresponding to the
Klein polynomials. 

In the instanton case, by contrast, we do not have such information
about $W$.  Since $W$ is identifiable with a space of Dirac
zero-modes, one can compute the character of $W$ as a
$\tG$-representation space using some equivariant index theory, but
the information coming from this is not particularly useful.  Instead
we exploit the fact that the representation theory of finite subgroups
of $SU(2)$ can be understood very explicitly.  Since every irreducible
representation of $\tG$ eventually appears in the standard
representations of $SU(2)$, the two approaches are closely related in
principle, though this may be rather cumbersome in practice.

We shall now, therefore, describe the irreducible representations of
the binary icosahedral group $\tG$ following John McKay's famous
paper\cite{MK}.  McKay observed that a convenient picture of the
representation theory of $\tG$ is given by the extended $E_8$ Dynkin
diagram:
$$
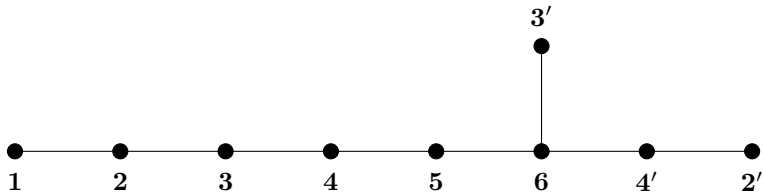
\epsfbox{e8.eps}
$$
Each node  stands for an irreducible representation of $\tG$. Here
$\one,\ldots,\six$ arise simply by restriction of the corresponding
representations of $SU(2)$ to $\tG$. The other representations $\twp,
\thp, \frp$, of dimensions $2,3,4$ respectively, will be described
shortly. First we shall explain the role of the edges in the diagram:
the node $\alpha$ is joined to $\beta$ if and only if $\alpha$ is a
summand in the decomposition of $\beta \otimes \two$ into irreducible
representations of $\tG$.  In general, this would lead to a directed
graph, but for subgroups of $SU(2)$, $\alpha$ occurs in
$\beta\otimes\two$ iff $\beta$ occurs in $\alpha\otimes
\two$. Moreover, the multiplicity is always $0$ or $1$.  Thus we read
off, for example $\six\otimes \two = \thp\oplus\frp\oplus\fv$. Since
moreover as $SU(2)$-representations, $\six\otimes \two =
\fv\oplus\sev$, we find that
$$
{\sev} = {\thp}\oplus{\frp}\mbox{ as representations of }\tG.
$$
This decomposition corresponds to the existence of the icosahedral
Klein polynomial of degree 12. Indeed the decomposition is equivalent
to the existence of a non-trivial $\tG$-map $\sev\to\sev$, or equally
to an invariant element in $\sev\otimes\sev$. Now the latter contains
the representation ${\bf 13}$ which contains the above-mentioned Klein
polynomial.  

Notice also that $\frp = \twp\otimes\two$. 

In view of this last observation, it remains to describe $\twp$ and
$\thp$.  In the character table of $\tG$, these two representations
look exactly like $\two$ and $\thr$, but with the sign of $\sqrt{5}$
changed. As was pointed out to the first author by J\o rgen Tornehave,
this extends to the representations, in the following sense. Identify
$\tG$ with a finite subset of points of $\HH$ such that all the
coordinates of these points lie in $\Q(\sqrt{5})$.  In other words, for
every $g\in \tG$, the coefficients of $i$, $j$, $k$, (as well as the
real part) are each of the form $a +b\sqrt{5}$, where $a$ and $b$ are
rationals. (We shall show how to do this in a moment.)  Consider the
`conjugation' on $\Q(\sqrt{5})$, $\alpha\mapsto \alpha'$ which takes
$a + b\sqrt{5}$ to $a -b\sqrt{5}$. Then $\twp$ is the representation
$x \mapsto g'x$, $\thp$ is the representation $x \mapsto g'x(g')\iv$
(on pure imaginary $x$) and $\frp$ is the representation $x\mapsto
g'xg\iv$. This recipe defines a representation because of the property
$\alpha'\beta' = (\alpha\beta)'$ for any $\alpha,\beta\in
\Q(\sqrt{5})$ and it is clear that the character of the `primed'
representation is obtained by changing the sign of $\sqrt{5}$, just as
required.  It so happens that only  the primed versions of $\two$
and of $\thr$ are new representations. Remark
that despite the notation, $\frp$ is not the primed version of
$\fr$; we hope that no confusion will result from this notation.

In this discussion note that $\thp$ and $\frp$ are real
representations: they act on $\Iim(\HH)$ or $\HH$, viewed as $3$ and
$4$-dimensional real vector spaces. By contrast, $\twp$ is essentially
complex; it is not the complexification of any real representation. We
point out further that $\frp$ has the following alternative
description.  A 5-dimensional representation of $\tG$ arises through
the identification of $G$ with the alternating group $A_5$ and letting
this act by permutations of the coordinates in $\R^5$. This is the sum
of a trivial representation and an irreducible $4$-dimensional
representation on $V\subset\R^5$, $V= \{y\in \R^5: y_1 + y_2 + y_3 +
y_4 + y_5=0\}$. This representation on $V$ is isomorphic to $\frp$. 

Coxeter, in \cite{RCP}, gives a description of the binary icosahedral
group (his notation is $\langle 5,3,2\rangle$) which has the property
mentioned above, namely that all coordinates lie in $\Q(\sqrt{5})$.  
For this, the icosahedron is oriented in such a way that the
coordinate axes in $\R^3$ pass through edge-midpoints.  In particular,
the half-turns about the coordinate axes lie in $G$. In terms of
quaternions, these half-turns correspond to $i$, $j$, $k$.  Then $\tG$
is generated by these together with one other element of order two
such as $U_2 = -(i + \tau j - \tau^{-1}k)/2$.  Setting
\be\label{nicg}
U_1 = j,\;U_2 = -(i + \tau j - \tau^{-1}k)/2,\;U_3= i\ee
one gets a set of generators of $\tG$.  In terms of these, the
five-fold rotation is given by $A = U_1U_2$, the three-fold rotation
by $B= U_2U_3$ and the two-fold rotation by $C=U_3U_1$ and
$$
A^5 = B^3 = C^2 = ABC
$$
and this equals $-1$ in $\tG$ but $1$ when projected to
$G$. (\cite{RCP}, 
p78, eqn (7.54) and p.69 eqn (6.65)).   Here $\tau = (\sqrt{5} + 1)/2$
is the golden ratio.

We can now write explicitly the action of the generators $U_1,U_2,U_3$
in the irreducible representations of $\tG$.  We shall confine
ourselves to the cases needed in this paper.  Where necessary, we
shall denote by $\rho_{\alpha}$ the action of $\tG$ in the
representation corresponding to $\alpha$. 
The following identities involving $\tau$ will be
used without comment in what follows:
$$
\tau^{-1} = (\sqrt{5} -1)/2,\; \tau' = -\tau\iv,\; \tau -\tau\iv =
1,\;\tau + \tau\iv = \sqrt{5},\; \tau^2 +\tau^{-2} =3.
$$ 
\begin{itemize}
\item  Thinking of $\two$ and $\twp$ as quaternionic 1-dimensional
representations, we have
$$
\rho_2(i) = \rho_{2'}(i) = i, \rho_2(j) = \rho_{2'}(j) = j,
\rho_2(k) =  \rho_{2'}(k) = k,
$$
and
$$
\rho_2(U_2) = -\frac{1}{2}(i + \tau j - \tau^{-1}k),\;
\rho_{2'}(U_2) = -\frac{1}{2}(i - \tau\iv j + \tau k).
$$

\item In terms of quaternions, $\thr$ is obtained by letting $\tG$ act
by conjugation on the imaginary quaternions $\Iim(\HH)$. 
Identifying $\Iim(\HH)$ with $\R^3$ via the coordinates $a_1i +
a_2j + a_2k$, we find 
$$
\rho_3(i) = \rho_{3'}(i)=\pmatrix{1 & 0 & 0\cr 0 & -1 & 0 \cr 0 & 0 & -1\cr},
\rho_3(j) = \rho_{3'}(j)=\pmatrix{-1 & 0 & 0\cr 0 & 1 & 0 \cr 0 & 0 & -1\cr},
$$
while
$$\rho_3(U_2) = -\frac{1}{2}
\pmatrix{ 1 & -\tau & \tau\iv\cr -\tau & -\tau\iv & 1\cr
\tau\iv & 1 & \tau\cr},\;
\rho_{3'}(U_2) = -\frac{1}{2}
\pmatrix{ 1 & \tau\iv & -\tau\cr \tau\iv & \tau & 1\cr
-\tau & 1 & -\tau\iv\cr}.$$
\item Identifying $\HH$ with $\R^4$ via the coordinates $a_0 + a_1i +
a_2j + a_3k$, we obtain the action in the representation $\frp$:
$$
\rho_{4'}(i) = \pmatrix{1 & 0 & 0 & 0\cr
                       0 & 1 & 0 & 0\cr
                       0 & 0 & -1 & 0\cr
                       0 & 0 & 0 & -1\cr},
\rho_{4'}(j) = \pmatrix{1 & 0 & 0 & 0\cr
                       0 & -1 & 0 & 0\cr
                       0 & 0 & 1 & 0\cr
                       0 & 0 & 0 & -1\cr},
$$
and
$$
\rho_{4'}(U_2) = \frac{1}{4}\pmatrix{-1 &  \sqrt{5} &- \sqrt{5} & -\sqrt{5}\cr
\sqrt{5} & 3 & 1 & 1\cr
-\sqrt{5} & 1 & -1 & 3\cr -\sqrt{5} & 1 & 3 & -1\cr}.
$$
\end{itemize}

\subsection{The ADHM description of an icosahedral 7-instanton}

To construct the ADHM data for an icosahedral instanton of charge $n$, we must
choose a real representation of $\tG$ of dimension $n$ (the space $W$)
and a $1$-dimensional quaternionic representation (the space
$E_\infty$). Having done so, we write down the most general $\tG$-invariant
maps and attempt to use them to solve the ADHM constraints.  

{\bf Notational remark}  In the rest of this paper, we shall
use the term `$\tG$-map' to denote any linear map between
representation spaces of $\tG$ that intertwines the $\tG$-actions.  We
shall also often use Schur's lemma without comment.

Following the monopole situation, we suppose $W = {\sev}= \thp\oplus
\frp$. From McKay's correspondence we have
$$
\thp\otimes \two = \six,\;\frp\otimes\two = \six\oplus\twp.
$$
Since $L$ is required to be a $\tG$-map from $W\otimes\two$ into
$E_\infty$, there is only one possibility for $E_\infty$ which allows
$L\not=0$: the only $2$-dimensional representation that occurs in
$W\otimes \two$ is $\twp$. Hence we take
\be\label{repan}
W = {\thp}\oplus {\frp},\; E_\infty = \twp.
\ee
The quaternionic matrix $M$ is naturally viewed as a map $W \to
W\otimes \two\otimes \two$. The McKay correspondence can be used also
to calculate the space of such $\tG$-maps. Now $M$ breaks up
naturally into a real component and a pure imaginary component,
corresponding to $\two\otimes\two = \one\oplus \thr$. The real
component of $M$ must be a multiple of the identity on the irreducible
summands $\thp$ and $\frp$, yielding two free parameters. The pure
imaginary part of $M$ gives maps $\thp \to \thp\otimes \thr$, $\thp
\to \frp\otimes \thr$, and $\frp \to \frp\otimes\thr$. From the McKay
correspondence, there are no non-zero maps $\thp \to \thp\otimes
\thr$, but there is in each case a one-dimensional space of maps 
$\thp \to \frp\otimes \thr$ and $\frp \to \frp\otimes\thr$. 

To sum up, we have the following parameter-count for $\tG$-symmetric ADHM
data: one in the component of $L$ that maps $\frp\otimes\two \to
\twp$; two from the real part of $M$; and three from the imaginary
part of $M$.  In the following lemma we compute explicitly these
invariant maps.  Although \maple has been used to assist with these
computations, they are quite straightforward to do if one first imposes
invariance under the group $\tK\subset \tG$ generated by
$i,j,k$.  Using the above formulae,
\begin{itemize}
\item The most general $\tG$-map $\frp\otimes\two \to \twp$
is given by any real multiple of the row-matrix $l=(1,\,i,\,j,\,k)$.

\item The most general $\tG$-map $\thp \to \frp\otimes\thr$ is given
by any real multiple of
$$
B = \pmatrix{i & j & k\cr
              0 & \tau k & \tau\iv j\cr
              \tau\iv k & 0 & \tau i\cr
              \tau j & \tau\iv i & 0\cr}.
$$
\end{itemize}
More specifically, this means that for each $g\in \tG$, we have
$$
g'\,l \,\rho_{4'}(g)\iv g\iv = l,
$$
and
$$
\rho_{4'}(g)g\, B\, \rho_{3'}(g)\iv g\iv = B.
$$
Taking the quaternionic conjugate of this,
$$
\rho_{3'}(g)g \,B^\dagger\, \rho_{4'}(g)\iv g\iv = B^\dagger.
$$
It follows that 
$$BB^\dagger = \pmatrix{3 & -i & -j & -k\cr i & 3 & k & -j\cr
j & -k & 3 & i\cr k & j & -i & 3\cr}
$$
satisfies $\rho_{4'}(g)g\, B
B^\dagger \,\rho_{4'}(g)\iv g\iv = BB^\dagger $, so its imaginary part
gives the 
unique $\tG$-map $\frp\otimes \thr \to \frp$. Since this is skew
symmetric, it cannot be used as a diagonal block in the matrix $M$
from the ADHM data, since $M$ is required to be symmetric. It follows
that the most general $\tG$-invariant ADHM data, with $W =
\thp\oplus\frp$, is given by
\be\label{invf}
\hat{M} = \pmatrix{ al & 0\cr b\identity_4 & cB\cr
-cB^\dagger & d\identity_3\cr}
\ee
where $a,b,c,d$ are real numbers and we have used the fact that $B$ is
pure imaginary to write $B^t = -B^\dagger$. One computes
$$
B^\dagger B =  4\identity_3,\;\; l^\dagger l= 4 \identity_4 - B
B^\dagger 
$$
so that
$$
\hat M^\dagger\hat M =
\pmatrix{ (4a^2+b^2)\identity_4  + (c^2-a^2)BB^\dagger & c(b-d)B\cr
c(b-d)B^\dagger & (4c^2 + d^2)\identity_3\cr}.
$$
Thus the ADHM constrainsts are satisfied iff $a^2 = c^2$ and
$c(b-d)=0$. If $c=0$, then $a=0$ and the top row $L$  of $\hat M$ is
identically zero. This yields a singular instanton and hence is not
allowable. Hence $c\not=0$, $b=d$ and $a = \pm c$. Since, moreover,
$$
\pmatrix{1 & 0 & 0\cr 0 & \identity_4 & 0\cr
0 & 0 & - \identity_3\cr} 
\pmatrix{ a l & 0\cr b\identity_4 & -aB\cr
aB^\dagger & b \identity_3\cr}
\pmatrix{\identity_4 & 0\cr
0 & - \identity_3\cr} =
\pmatrix{ al & 0\cr b\identity_4 & aB\cr
-aB^\dagger & b\identity_3\cr}
$$
the two choices of sign lead to gauge-equivalent ADHM data and hence
to gauge-equivalent instantons. 

What we have found is a charge-7 icosahedral instanton that is unique
up to the obvious freedom to translate along the $x_4$ axis (the
parameter $b$) and overall scale ($(a,b)\mapsto (\lambda a,\lambda
b)$).  If we centre the instanton at the origin of $\R^4$ then the
ADHM data are given by any real multiple of
\be\label{icinst}
\pmatrix{ 1&i&j&k&0&0 & 0\cr
0&0& 0 & 0 & i&j&k\cr
0&0&0&0&0&\tau k & \tau\iv j\cr
0&0&0&0&\tau\iv k &0&\tau i\cr
0&0&0&0&\tau j&\tau\iv i &0 \cr
i& 0&\tau\iv k & \tau j & 0 & 0 & 0\cr
j& \tau k & 0 &\tau\iv i&0&0&0  \cr 
k & \tau\iv j & \tau i & 0&0&0&0\cr}
\ee
A \maple calculation shows that $\Delta(x)^\dagger\Delta(x)$ is
invertible for every $x$, so that this is a non-singular instanton.

\subsection{The associated Skyrmion}

Having constructed the ADHM data for an icosahedral instanton we now
wish to make use of this to compute a Skyrme field. Recall the
proposal of Atiyah and Manton\cite{AM} which results in the following
explicit prescription for the Skyrme field \be U({\bf x})={\cal
P}\exp\bigg(-\int_{-\infty}^{+\infty} A_4({\bf x},x_4)\ dx_4\bigg).
\ee Here $U({\bf x})$ is the $SU(2)$-valued Skyrme field in $\R^3$,
${\cal P}$ denotes path ordering and $A_\mu$ is the gauge potential of
a Yang-Mills instanton field in $\R^4.$

For an instanton of charge $n$ this holonomy produces a Skyrme field with baryon number $n.$
Although this procedure does not give exact solutions to the Skyrme model it does give
fields which are good approximations to important Skyrmion configurations, 
in the sense of having not only the correct symmetries but also energies which are only a few
 percent above those of the numerically known Skyrmion solutions.
For example, the 1-instanton generates a hedgehog Skyrme field and by adjusting the scale
of the instanton, which may be regarded as a free parameter in the approximation, it is possible
to obtain an approximation whose energy is only 1\% above that of the true solution (which is known
only numerically).

The minimal energy Skyrmions of charge two, three and four have axial, tetrahedral
and cubic symmetry respectively\cite{BTC}. In each of these cases instantons have been found with the correct
symmetries, so that computing their holonomies produces  Skyrme fields which are good approximations
to these Skyrmions\cite{AM,LM}. It should be noted that in these multi-instanton examples the holonomy
can not be computed analytically and therefore numerical methods must be employed. Specifically, if
$\widetilde U({\bf x},x_4)$ denotes the solution of the matrix ordinary differential equation
\be
\partial_4\widetilde U=-A_4\widetilde U
\label{ode}
\ee
with the initial condition $\widetilde U({\bf x},-\infty)=\identity_2,$ then 
 $U({\bf x})=\widetilde U({\bf x},\infty).$ The above set of ordinary differential equations, with ${\bf x}$
regarded as a parameter, need to be solved numerically, for which we employ a standard Runge-Kutta method.
In principle, since we have the ADHM data explicitly, it is possible to obtain an exact analytic expression
for $A_4$ by performing some quaternionic linear algebra. However, this is a non-trivial computation, even with the
use of a symbolic computer algebra package, and is of little practical use. Furthermore, since we are 
employing a numerical algorithm to solve equation (\ref{ode}) then it is sensible to compute
$A_4$ numerically also. To achieve this, note that equation (\ref{construction}) states that the vector
$N(x)$ is orthogonal to each of the columns of the matrix $\Delta(x).$ Thus we can compute $N(x)$ using
a quaternionic Gram-Schmidt orthonormalization process and hence $A_4(x)$ by using a finite difference
approximation to equation (\ref{afromn}).

Applying our numerical scheme to the ADHM data of the icosahedral 7-instanton derived above we obtain
a Skyrme field whose baryon density isosurface is displayed in Figure 1.11. This surface, which looks identical
to that obtained from full field simulations of the Skyrme model\cite{BS2}, resembles a dodecahedron, with the 
baryon density being maximal at the vertices of the dodecahedron and holes at the centre of each face.
We have not attempted to find the instanton scale at which the energy of this Skyrme field is minimized since
to perform an accurate calculation would require a more substantial amount of computing time.
The important point is that we have demonstrated that an icosahedrally symmetric 7-instanton exists whose
holonomy produces a good approximation to the minimal energy 7-Skyrmion.

\section{Seven Skyrmion scattering}
\news\ \ \ \ \ \
The aim of this section is to obtain a family of instantons which describes a seven Skyrmion scattering
process in which seven well separated Skyrmions merge to form the dodecahedral 7-Skyrmion. 
In order to understand this scattering it is useful to first consider the analogous situation for monopoles,
since from numerical simulations it appears that many Skyrmion scattering events are remarkably
 similar to known monopole scatterings\cite{BS1}.

A dodecahedral 7-monopole exists\cite{HS2} whose energy density isosurface looks very similar to
the baryon density isosurface shown in Figure 1.11. In fact it was known earlier that a 
 tetrahedral 3-monopole and a cubic 4-monopole also exist\cite{HMM} and again they resemble the
corresponding Skyrmions\cite{HS1}. In studying symmetric monopoles it is useful to consider rational
maps, which we outline below.

The $n$-monopole moduli space is diffeomorphic to the space of
degree $n$ rational maps between Riemann spheres, with the equivalence relation that two maps that are
 equal after a rotation of the target sphere are identified. The rational map arises from the
monopole as the scattering data along a half-line emanating from a chosen origin\cite{Ja}. Explicitly,
let $z$ be a point on the Riemann sphere and consider Hitchin's equation
\be (D_r-\mbox{i}\Phi)s=0
\label{hitchin}
\ee
for the two-component field $s$, along the radial half-line through the point $z.$ Here $D_r$ is
the covariant derivative in the radial direction and $\Phi$ is the Higgs field. Up to a constant
multiple, there is a unique solution $s=(s_1,s_2)^t$ which decays as $r\rightarrow\infty.$
Let $R$ be the ratio of the components of this solution evaluated at the origin, that is,
$R=\frac{s_1}{s_2}\vert_{r=0}.$ Now consider how $R$ varies as we choose a new direction for
the half-line by changing the value of $z.$ Then, as proved by Jarvis\cite{Ja}, $R$ is a holomorphic
function of $z$ of degree $n$, where $n$ is the charge of the monopole fields occuring in (\ref{hitchin}).
The effect of a gauge transformation is to transform $R$ by an $SU(2)$ M\"obius transformation, and
after taking this equivalence into account, there is a one-to-one correspondence between $n$-monopoles
and rational maps $R(z)$ of degree $n.$

If a map $R(z)$, of degree $n$, is $G$-invariant (up to M\"obius transformations) then there
is an $n$-monopole with symmetry $G$, and vice versa. In ref.\cite{HMS} many symmetric maps are
presented but the one of relevance here is the following degree seven map which arises after
the imposition of the symmetry $T_h$
\be
R(z)=\frac{bz^6-7z^4-bz^2-1}{z(z^6+bz^4+7z^2-b)}.
\label{g7}
\ee 
$T_h$ is the group of rotations of a tetrahedron extended by inversion symmetry and, after
a choice of orientation, the above one-parameter family, with $b$ real, gives all such maps.
Since this one-parameter family is the fixed point set of a group action then it is a geodesic
in the 7-monopole moduli space. Using the geodesic approximation\cite{Ma1} this family describes
a low energy seven monopole scattering process as $b$ varies along the real line from $-\infty$
to $\infty.$ Changing the sign of $b$ can be undone with a M\"obius transformation plus the
replacement $z\mapsto \mbox{i}z$, which corresponds to a rotation by $90^\circ$ about the $x_1$-axis.
If $b=\pm 7/\sqrt{5}$ then the map has icosahedral symmetry, and represents the dodecahedral
7-monopole and its dual, whereas at $b=0$, which is the midpoint of the scattering process,
 the map has cubic symmetry. In the limit as $b\rightarrow\infty$ the map degenerates
to $R(z)=z$, which represents a single monopole at the origin, the other six monopoles
having moved off to infinity along the Cartesian axes.

In summary this geodesic models a scattering event where six monopoles, moving in along
the Cartesian axes, merge with a single monopole at the origin to form first a dodecahedron
and then a cube, after which the process reverses but with a $90^\circ$ rotation.
The purpose of the remainder of this section is to present a one-parameter family of
Skyrme fields which describe a similar scattering of Skyrmions. At this point it is
important to note that although an ansatz exists for Skyrme fields in terms of rational maps\cite{HMS},
this approximation only works for Skyrme fields which have a shell-like structure. Thus, for example,
using the rational map (\ref{g7}) with $b=7/\sqrt{5}$ gives a good approximation to the
dodecahedral 7-Skyrmion, but the rational map ansatz breaks down for large $b$ and can not be used
to describe well separated Skyrmions. Thus we need to turn to the instanton approximation to
attempt to produce Skyrme fields which describe this process.

The upshot of the above discussion is that we now want to consider 7-instantons with symmetry
$T_h.$ 

\subsection{Tetrahedral deformations of the icosahedral instanton}

We have described the general framework for the construction of
symmetric ADHM data. What we seek now is the most general family of
tetrahedral $7$-instantons that contains the icosahedral
$7$-instanton.  Denote by $\tT$ the binary tetrahedral group, i.e. the
double cover in $SU(2)$ of the rotation group of the tetrahedron. To
study tetrahedral deformations, we realize $\tT$ as a subgroup of
$\tG$ generated by $i,j,k$ and the three-fold rotation
$(1+i+j+k)/2$. (We continue to take $\tG$ to be generated by the $U_i$
of (\ref{nicg}).) Conjugation by this quaternion gives a
$120^\circ$-rotation about the axis in the direction $(1,1,1)$ in $\R^3$.
Then the given representations of $\tG$ yield representations of $\tT$
and we can attempt to follow the procedure described above. It is
preferable first to  impose the inversion from the subgroup $T_h$,
however, since this forces the two diagonal blocks in $M$ to be zero.

To explain this, we must first show in what sense the data
(\ref{icinst}) are inversion symmetric.  For this we must again find
compensating gauge transformations $J_1$ and $J_2$ such that
$$
\Delta(-x) = J_1\Delta(x)J_2\iv.
$$
It is easy to see that the essentially unique choice for this, in the
case of the ADHM data (\ref{icinst}), is given by
$$
J_1 = \pmatrix{-1 & 0 & 0\cr 0 & \identity_4 & 0\cr 0 & 0 &
-\identity_3\cr},\;\;
J_2 = \pmatrix{-\identity_4 & 0\cr 0 & \identity_3\cr}.
$$
The reason for the simple form of these matrices is that the inversion
$x \mapsto -x$ is central and so must act by $\pm$ the identity in any
irreducible representation of $\tG$.

Now any ADHM data that is inversion-symmetric in this sense must have
the form
$$
\hat M = \pmatrix{v & 0\cr 0 & C \cr C^t & 0 \cr}
$$ 
relative to the usual block-decomposition. The
row-vector $v$ and $4\times 3$ matrix $C$ are constrained by having to
be symmetric under $\tT$, where this acts by restriction of the given
$\tG$-representations. Since $\two = \twp$ and $\thr =\thp$ as $\tT$
representations, we have that $\frp = \two\otimes \two =
\one\oplus\thr$ as  $\tT$-representations. One may check that $v$ is
invariant iff it has the form
$(\lambda_0,\,\lambda_1i,\,\lambda_1j,\,\lambda_1k)$ 
and that $C$ is invariant if it has the form
$$
C = \pmatrix{\lambda_2 i &\lambda_2  j &\lambda_2  k \cr
p & qk & rj \cr
rk & p & qi \cr
qj & ri & p\cr}
$$
where $\lambda_0,\lambda_1,\lambda_2,p,q,r$ are real numbers.  

The next task is to impose the ADHM constraints; these yield the
equations
$$
\Iim(v^\dagger v + \bar{C}C^t) = 0,\;\Iim(C^\dagger C) = 0.
$$
It is straightforward to show that these reduce to the three equations
$$
\lambda_0\lambda_1 = \lambda_2(p+q-r),\;
\lambda_1^2 = qr - p(q-r),\; \lambda_2^2 = qr + p(q-r).
$$
Solving for the $\lambda$'s in terms of $p,q,r$, we conclude that
the general $T_h$ invariant ADHM data are given by
\be\label{adhmp}
\hat M=\pmatrix{\lambda_0 & \lambda_1 i & \lambda_1 j &\lambda_1 k & 0 & 0 & 0\cr
         0 & 0 & 0 & 0 &\lambda_2 i &\lambda_2  j &\lambda_2  k \cr
         0 & 0 & 0 & 0 & p & qk & rj \cr
         0 & 0 & 0 & 0 & rk & p & qi \cr
         0 & 0 & 0 & 0 & qj & ri & p\cr
        \lambda_2  i & p & rk & qj & 0 & 0 & 0\cr
       \lambda_2   j & qk & p & ri & 0 & 0 & 0\cr
       \lambda_2   k & rj & qi & p & 0 & 0 & 0\cr}
\ee
where 
\be\label{adhmt}
\lambda_0=(p+q-r)\sqrt{\frac{qr+p(q-r)}{qr-p(q-r)}}, \
\lambda_1=\sqrt{qr-p(q-r)}, \ \lambda_2=\sqrt{qr+p(q-r)}.
\ee

Here we have three free parameters, $p,q,r$, but we require only a
one-parameter family for our application.  One of the free parameters
is an overall scale 
factor, which corresponds to the multiplication of $\hat M$ by a
scale, and this can only be determined by computing the energy of the
associated Skyrme field. We will therefore set this scale to one, as
it can be easily reintroduced later.  As $p,q,r$ are homogeneous
coordinates we can fix this scale by setting $r=q^{-1}$.

To get some insight into the interpretation of the parameters, let us
consider the special case $p=0$. Then from (\ref{adhmt}), $\lambda_0 =
\qd$, 
$\lambda_1=\lambda_2=1$; denote the corresponding matrix (\ref{adhmp})
by $\hat M(q)$. If
$q=\tau$, then $\lambda_0 =1$ and $\hat M(\tau)$ coincides with the
icosahedral matrix (\ref{icinst}). If $q$ is interpreted as
the exponential of a `time parameter' $t$, then $q$ runs from $0$ to
$\infty$ as $t$ runs from $-\infty$ to $\infty$, and time-reversal
is the transformation $q \to q\iv$. This transformation is equivalent
to a $90^\circ$ rotation about the $x_1$ axis. Indeed if we denote by
$\hat N(q)$ the ADHM data obtained from $\hat M(q)$ by replacing $q$
by $q\iv$, $i$ by $i$, $j$ by $k$ and  $k$ by $-j$, then we have
\be \label{co1}
\hat N(q) =\pmatrix{-1 & 0 & 0\cr
                      0 & R_1 & 0\cr 0 & 0 & R_2\cr}M(q)
\pmatrix{R_1 & 0 \cr 0 & R_2}\iv
\ee
if
\be \label{co2}
R_1 = \pmatrix{ 1 & 0 & 0 & 0\cr 0&-1&0&0\cr 0 & 0&0&-1\cr0&0&1&0\cr},\;
R_2 = \pmatrix{1 & 0 & 0\cr 0 & 0 &1\cr 0 &-1 &0\cr}.
\ee
We conclude that the value $q= \tau\iv$ is also a dodecahedral
instanton (obtained from (\ref{icinst}) by a $90^\circ$ rotation) and
that at $q=0$ we have an instanton symmetric under the symmetry group
of the cube (since this group is generated by the tetrahedral group
together with any $90^\circ$ rotation about one of the coordinate
axes).  

This one-parameter family thus has all the symmetry properties
expected of the seven Skyrmion scattering process.  However it
turns out that it does not have the correct asymptotic behaviour as
$q$ goes to $0$ or $\infty$.  This was verified both by numerical work and
asymptotic analysis. The problem can be traced to the fact that the
first entry of the top row $\qd$ blows up as $q\to 0,\infty$. 

Therefore we try to find a more general family with the same symmetry
properties but allow $p\not=0$ to improve the asymptotic behaviour. In
particular we want time-reversal to continue to correspond to  the
replacement $q \mapsto q\iv$.  Then it can be seen that if we make the
replacements
\be
q\mapsto q\iv, p\mapsto -p, i\mapsto i, j\mapsto k, k\mapsto -j
\ee
which, because of (\ref{adhmt}), result in 
\be
\lambda_0\mapsto-\lambda_0,
\lambda_1\mapsto\lambda_1,\lambda_2\mapsto\lambda_2,
\ee
equivalent ADHM data are obtained.  (The `compensating gauge
transformation' is as in (\ref{co1}) and (\ref{co2}).)   
Note that the fixed point set of the time reversal transformation, in other
words the midpoint of the scattering process, is given by $q=1, \ p=0$, which
indeed has cubic symmetry as it should.

These arguments show that if $p$ is any function of $q$ with
the properties
\be\label{prp}
p(q\iv) = -p(q),\;\;p(\tau)= 0 
\ee
then the corresponding family of ADHM matrices will have the correct
symmetry properties: the first condition gives that the data at $q$
and at $q\iv$ are equivalent up to a $90^\circ$-rotation; the second
ensures that the data at $\tau$ reduce to the icosahedral data (\ref{icinst}).

We are now left with finding the variable $p$ as a function of $q$;
this can not be determined from symmetry arguments alone. 
A simple function satisfying (\ref{prp}) is
given by \be p=-\frac{\qd(\qd^2-1)}{\qd^4+1}.
\label{nicep}
\ee 
The numerator of $p$ is the simplest function having the required
zeros and symmetry properties, whereas at this stage the only fact we
know about the denominator is that it must be a symmetric function of
$\qd$. However, as we shall now see, the form of the denominator is
highly constrained by examining the asymptotic limit of the
scattering process, where all seven instantons are well separated,
and this leads naturally to the given solution.

The asymptotic out state corresponds to the limit $q\rightarrow\infty$,
and in this limit the leading order behaviour of $p$ is
\be
p=-q^{-1}+O(q^{-5}).
\ee
Hence in this limit we find that
\be
\lambda_0=\frac{1}{\sqrt{2}}+O(q^{-2}),\
\lambda_1=\sqrt{2}+O(q^{-2}), \
\lambda_2=q^{-1}+O(q^{-3}).
\ee
Neglecting negative powers of $q$ we thus arrive at the asymptotic ADHM data
\be
\hat M_\infty=\pmatrix{1/\sqrt{2} & \sqrt{2} i &  \sqrt{2} j & \sqrt{2} k & 0 & 0 & 0\cr
         0 & 0 & 0 & 0 & 0 & 0 & 0 \cr
         0 & 0 & 0 & 0 & 0 & qk & 0 \cr
         0 & 0 & 0 & 0 & 0 & 0 & qi \cr
         0 & 0 & 0 & 0 & qj & 0 & 0\cr
         0 & 0 & 0 & qj & 0 & 0 & 0\cr
         0 & qk & 0 & 0 & 0 & 0 & 0\cr
         0 & 0 & qi & 0 & 0 & 0 & 0\cr}.
\ee
After a gauge transformation by the matrix 
\be
T = \frac{1}{\sqrt 2}\pmatrix{
      \sqrt{2} & 0 & 0 & 0 & 0 & 0 & 0\cr
      0 & 1 & 0 & 0 & 0 & 1 & 0\cr
      0 & 0 & 1 & 0 & 0 & 0 & 1 \cr
      0 & 0 & 0 & 1 & 1 & 0 & 0\cr
      0 & 0 & 0 & 1 & -1 & 0 & 0\cr
      0 & 1 & 0 & 0 & 0 & -1 & 0\cr
      0 & 0 & 1 & 0 & 0 & 0 & -1\cr}
\ee
we obtain
\be
\pmatrix{1 & 0\cr 0 & T\cr}\hat M_\infty T^{-1}=
\pmatrix{ 1 & i & j & k &
k & i & j\cr
0 & 0 & 0 & 0 & 0 & 0 & 0 \cr
0 & -qk & 0 & 0 & 0 & 0 & 0\cr
0 & 0 & -qi & 0 & 0 & 0 & 0\cr
0 & 0 & 0 & -qj & 0 & 0 & 0\cr
0 & 0 & 0 & 0  & qj & 0 & 0\cr
0 & 0 & 0 & 0  & 0 & qk & 0\cr
0 & 0 & 0 & 0  & 0 & 0 & qi \cr}.
\ee
This matrix is of the form identified by Christ, Stanton and Weinberg\cite{CSW} as representing well
separated instantons. The leading order terms lie on the diagonal of the square part of the matrix
and determine the instanton positions, giving one instanton at the origin and the other six on the
Cartesian axes at a distance $q$ from the origin. The terms of next order all lie on the top row of
the matrix and these give the scales and $SU(2)$ orientations of the instantons. The fact that all the entries
on the top row have unit length means that each instanton has scale one. Since the first entry in the top
row is one, then on computing the holonomy the Skyrmion at the origin will be in standard orientation,
whereas, for example, the fact that the second and sixth entries in the top row are $i$, means that the Skyrmions
located on the $x_3$-axis have an orientation which is obtained from the standard one by rotating the Skyrmion
by $180^\circ$ around the $x_1$-axis. These kinds of configurations, that is three collinear Skyrmions such that
the outer Skyrmions are rotated by $180^\circ$ about a line perpendicular to the line joining the inner Skyrmion,
are known to give attractive initial conditions for full field simulations\cite{BS1}. Thus we see that
our ADHM data has the correct asymptotic properties to produce
the configuration which we require. 

Now we address the possible freedom in the choice (\ref{nicep}) that
we have made. The main effect of changing the denominator in (\ref{nicep})
is to alter the scale of the instanton at the origin in the limit
in which the other six are far from it. The requirement that this scale
is finite as $q\rightarrow\infty$ determines that the leading term in the
denominator of $p$ must be $\qd^4$ with coefficient one. Only even powers
of $\qd$ are allowed by symmetry and it can be shown that
the coefficient of $\qd^2$ must be zero if the scale of the instanton at the
origin is to be not only finite but equal in value to the scale of the instantons
which are moving along the Cartesian axes. There remains the freedom to change
the value of the constant in the denominator of (\ref{nicep}) but clearly this
has little effect since it is only relevant for small values of $\qd$ and the
numerator contains $\qd$ as a factor.

Using this ADHM data for increasing values of $q$ we compute Skyrme fields
whose baryon density isosurfaces are shown in Figure 1. The various values
of $q$ corresponding to each figure are given in the table below.\\

\begin{center}
\begin{tabular}{|c|c|c|c|c|c|c|c|c|c|c|c|c|c|c|c|} \hline
Fig & 1 & 2 & 3 & 4 & 5 & 6 & 7 & 8 \\
\hline
q & 0.40 & 0.44 &0.48&0.50&0.62&0.77&0.87&1.00\\
\hline
Fig & 9 & 10 & 11 & 12 & 13 & 14 & 15 &  \\
\hline
q & 1.15 & 1.30 &1.62&2.00&2.10&2.25&2.50& \\
\hline
\end{tabular}
\end{center}
\begin{center}
Table 1. {\sl Parameter values for the scattering shown in Figure 1.}
\end{center}

\mbox{}From Figure 1 we see that indeed our family of instantons describes
the sought after scattering process. Figure 1.1 clearly shows the six
Skyrmions on the Cartesian axes and a Skyrmion at the origin. As the
Skyrmions approach, Figure 1.2, the one at the origin shrinks until it
disappears completely, Figure 1.3. The Skyrmions then merge until the
dodecahedron is formed, Figure 1.5, after which the configuration
twists until it turns into a cube, Figure 1.8.  This process is then
reversed but accompanied by a $90^\circ$ rotation around the
$x_1$-axis, so that for example the dual dodecahedron is formed,
Figure 1.11, and the Skyrmions finally separate again along the
Cartesian axes, Figure 1.15.

An important difference between monopoles and Skyrmions is that monopoles are BPS solitons,
and therefore all configurations of the same charge have equal energy, whereas for Skyrmions
this is certainly not the case and the potential energy of different configurations is an
important factor in considering the dynamics. We have not computed the scale factor of each
configuration to minimize the energy of our Skyrme fields since, although this could be done
if required, it would involve a substantial amount of computing time. However, we have ensured
that the scale factor is the order of unity for all $q$, so that we get an accurate representation
of the baryon density isosurfaces. Qualitatively we know that for all the family of Skyrme fields
the energy per Skyrmion is less than it is for seven well separated Skyrmions, and that the
minimum energy configurations are the two dodecahedra. From numerical simulations\cite{BS3}
it is known that the energy of the cubic 7-Skyrmion is above that of the dodecahedron
but it is still substantially less than that of seven well separated Skyrmions.
Thus the true dynamical evolution depends upon the initial speeds of the incoming Skyrmions,
together with the amount of energy lost through radiation as the process evolves. However,
it is expected that if the incoming speeds are great enough then the whole scattering process
displayed in Figure 1 would take place. Radiation effects will mean that for most speeds the
incoming Skyrmions will eventually get trapped at one of the dodecahedra, and perhaps if
the Skyrmions are initially static then only the first portion of the scattering process
will occur and the cube may never be formed, but this depends upon the amount of radiation
generated. By performing full field simulations, using the numerical code described
in \cite{BS1,BS3}, with initial conditions given by the instanton generated Skyrme field,
it has been verified that the true dynamical evolution does follow the sequence described
above and the family of instanton generated Skyrme fields provides an accurate approximation to the 
Skyrmion scattering process.

In current approaches to the quantization of Skyrmions a first step is to examine
the vibrational modes of the minimal energy Skyrmion at each charge\cite{BBT}. Thus for charge
seven it is the vibrational modes of the dodecahedral Skyrmion which need to be studied and
clearly one mode is the tetrahedral deformation we have displayed. Thus if the amplitude of the
deformation is sufficient then one of the important vibrational modes will be the one considered
here where the dodecahedron deforms to its dual via a cube.

\section{Conclusion}
\news\ \ \ \ \ \
We have used the ADHM construction to present a charge seven instanton with icosahedral
symmetry whose holonomy generates a Skyrme field which approximates the minimial energy 
dodecahedral 7-Skyrmion.
Furthermore, by imposing tetrahedral symmetry we have found a family of ADHM data which
we used to generate Skyrme fields that model a seven Skyrmion scattering process that results
in the formation of the dodecahedral 7-Skyrmion. 

There are a number of other highly symmetric Skyrmions, such as the icosahedrally symmetric
charge seventeen Skyrmion which resembles a buckyball\cite{HMS}, and the methods used here
could also be applied to construct the ADHM data of the predicted corresponding instantons.
In particular it would be interesting if the instanton approach led to an understanding of the
formation of the buckyball 17-Skyrmion from individual Skyrmions.\\

\section*{Acknowledgements}
\news\ \ \ \ \ \
Many thanks to Nick Manton and Erick Weinberg for useful
discussions; to Iain Rendall for producing the Dynkin diagram in
\S\ref{repthy} and to Toby Bailey and Elmer Rees for assistance with
\maple.  We also acknowledge the EPSRC for Advanced Fellowships 
and the grant GR/L88320.\\

\section*{Figure Caption}  

\noindent Fig.~1. Baryon density isosurfaces for a family of charge seven Skyrme fields obtained from 
7-instantons with tetrahedral symmetry.


\newpage

\begin{thebibliography}{xx}

\bibitem{ADHM} M.F. Atiyah, N.J. Hitchin, V.G. Drinfeld and Yu.I. Manin, Phys. Lett. A 65,
185 (1978).

\bibitem{AM} M.F. Atiyah and N.S. Manton, Phys. Lett. B 222, 438 (1989); Commun. Math.
 Phys. 153, 391 (1993).

\bibitem{BBT} C. Barnes, W.K. Baskerville and N. Turok, Phys. Rev. Lett. 79, 367 (1997);
 Phys. Lett. B 411, 180 (1997).

\bibitem{BS2} R.A. Battye and P.M. Sutcliffe, Phys. Rev. Lett. 79, 363 (1997).

\bibitem{BS1} R.A. Battye and P.M. Sutcliffe, Phys. Lett. B 391, 150 (1997).

\bibitem{BS3} R.A. Battye and P.M. Sutcliffe, preprint DAMTP-1998-108.

\bibitem{BTC} E. Braaten, S. Townsend and L. Carson, Phys. Lett. B 235, 147 (1990).

\bibitem{CSW} N.H. Christ, E.J. Weinberg and N.K. Stanton, Phys. Rev. D18, 2013 (1978).

\bibitem{RCP} H.S.M.Coxeter, `Regular Complex Polytopes' CUP (1974).
\bibitem{HMM} N.J. Hitchin, N.S. Manton and M.K. Murray, Nonlinearity, 8, 661 (1995).

\bibitem{HMS} C.J. Houghton, N.S. Manton and P.M. Sutcliffe,  Nucl. Phys. B 510, 507 (1998).

\bibitem{HS2} C.J. Houghton and P.M. Sutcliffe, Nonlinearity 9, 385 (1996).

\bibitem{HS1} C.J. Houghton and P.M. Sutcliffe, Commun. Math. Phys. 180, 343 (1996).

\bibitem{Ja} S. Jarvis, \lq{\sl A rational map for Euclidean monopoles via
radial scattering}\rq, Oxford preprint (1996).

\bibitem{LM} R.A. Leese and N.S. Manton, Nucl. Phys. A 572, 675 (1994).

\bibitem{Ma7} N.S. Manton, Phys. Rev. Lett. 60, 1916 (1988).

\bibitem{Ma1} N.S. Manton, Phys. Lett. B 110, 54 (1982).

\bibitem{MK} J. McKay, {\em in} Proc. Sympos. Pure Math. AMS {\bf 37},
183, (1980).
\end{thebibliography}
\end{document}